\documentclass[lineno]{article}

\usepackage[utf8]{inputenc}
\usepackage{graphicx}
\usepackage{amsfonts}
\usepackage[square,numbers]{natbib}
\usepackage{subcaption}
\usepackage{url,lineno}
\usepackage{tikz}
\usepackage{float}
\usepackage[normalem]{ulem}
\usepackage[labelfont=bf, font=scriptsize]{caption}
\usepackage{placeins}
\usepackage{authblk}
\usepackage[a4paper, total={6.3in, 10in}]{geometry}
\usepackage[colorlinks=true, citecolor=blue, filecolor=blue, linkcolor=black]{hyperref}
\usepackage{bbm}
\usepackage{color}
\usepackage{epstopdf}
\usepackage{stmaryrd}
\usepackage{comment}
\usepackage{multicol}
\usepackage{amsmath,amssymb} 
\usepackage{color}
\usepackage{mathtools}
\usepackage[T1]{fontenc}    
\usepackage{url}            
\usepackage{booktabs}       
\usepackage{amsfonts}       
\usepackage{nicefrac}       
\usepackage{microtype}      
\usepackage{lipsum}
\usepackage{xcolor,colortbl}
\usepackage{amsmath}
\usepackage[square,numbers]{natbib}
\usepackage{lastpage,fancyhdr}

\newcommand{\be}{\begin{equation}}
\newcommand{\ee}{\end{equation}}
\newcommand{\bea}{\begin{eqnarray}}
\newcommand{\eea}{\end{eqnarray}}

\definecolor{lime}{HTML}{A6CE39}
\DeclareRobustCommand{\orcidicon}{%
	\begin{tikzpicture}
	\draw[lime, fill=lime] (0,0) 
	circle [radius=0.16] 
	node[white] {{\fontfamily{qag}\selectfont \tiny ID}};
	\draw[white, fill=white] (-0.0625,0.095) 
	circle [radius=0.007];
	\end{tikzpicture}
	\hspace{-2mm}
}

\foreach \x in {A, ..., Z}{%
	\expandafter\xdef\csname orcid\x\endcsname{\noexpand\href{https://orcid.org/\csname orcidauthor\x\endcsname}{\noexpand\orcidicon}}
}

\begin{document}  

\title{\textbf{Critical fragility in socio-technical systems}}

\author[1,2,3]{Jos\'e Moran\orcidA{}}
\affil[1]{Macrocosm, Inc., Brooklyn, NY}
\affil[2]{Mathematical Institute and Institute for New Economic Thinking at the Oxford Martin School, University of Oxford, Oxford, United Kingdom}
\affil[3]{Complexity Science Hub, Vienna,  Austria}
\author[4,5]{Frank P. Pijpers\orcidB{}}
\affil[4]{Statistics Netherlands, The Hague}
\affil[5]{Korteweg-de Vries Institute for Mathematics, University of Amsterdam, The Netherlands}
\author[6,7,8]{Utz Weitzel\orcidC{}}
\affil[6]{School of Business and Economics, Vrije Universiteit Amsterdam, Amsterdam, The Netherlands}
\affil[7] {Tinbergen Institute, Amsterdam, The Netherlands}
\affil[8] {Nijmegen School of Management, Radboud University, Nijmegen, The Netherlands}
\author[9,10]{Jean-Philippe Bouchaud\orcidD{}}
\affil[9]{Capital Fund Management, Paris, France}
\affil[10]{X-CFM Chair of EconophysiX, Ecole polytechnique, Palaiseau, France}
\author[11,12]{Debabrata Panja\orcidE{}\footnote{Corresponding author: d.panja@uu.nl}}
\affil[11]{Department of Information and Computing Sciences, Utrecht}
\affil[12]{Centre for Complex Systems Studies, Utrecht University, The Netherlands}



\maketitle 

\begin{abstract}
Socio-technical systems, where technological and human elements interact in a goal-oriented manner, provide important functional support to our societies. Here we draw attention to the underappreciated concept of timeliness --- i.e., system elements being available at the right place at the right time --- that has been ubiquitously and integrally adopted as a quality standard in the \textit{modus operandi\/} of socio-technical systems. We point out that a variety of incentives, often reinforced by competitive pressures, prompt system operators to myopically optimize for efficiencies, running the risk of inadvertently taking timeliness to the limit of its operational performance,  correspondingly making the system critically fragile to perturbations by pushing the entire system towards the proverbial `edge of a cliff'.  Invoking a stylized model for operational delays, we identify the limiting operational performance of timeliness, as a true critical point, where the smallest of perturbations can lead to a systemic collapse. Specifically for firm-to-firm production networks, we suggest that the proximity to \textit{critical fragility\/} is an important ingredient for understanding the fundamental ``excess volatility puzzle'' in economics. Further, in generality for optimizing socio-technical systems, we propose that critical fragility is a crucial aspect in managing the trade-off between efficiency and robustness.
\end{abstract}

\vspace{5mm}
S ocio-technical systems (STSs) are complex systems where human elements (individuals, groups, and larger organisations), technology and infrastructure combine, and interact, in a goal-oriented manner. Their functionalities require designed or planned interactions among the system elements --- humans and technology --- that are often spread across geographical space. The pathways for these interactions are designed and planned with the aim of providing operational stability of STSs, and they are embedded within technological infrastructures \cite{Kaur2022}. Playing crucial roles in health services, transport, communications, energy provision, food supply, and, more generally, in the coordinated production of goods and services, they make our societies function. STSs exist at many different levels, from niche systems like neighborhood garbage disposal, to intermediate systems such as regional/national waste management, reaching up to systems of systems, e.g., global climate coordination in a world economy. 

\vspace{1mm}
\noindent  In spite of the design of the STSs with the intention of providing stable operations, STSs display the hallmarks of fragility, where the emergence of non-trivial dynamical instabilities are commonplace \cite{DeWeck2011}. Minor and/or geographically local events can cascade and spread to lead to system-wide disruptions, including a collapse of the whole system. Examples include (i) the grounding of an entire airline [e.g., Southwest Airlines in April 2023 \cite{Newsweek2023}]; (ii) the cancellation of all train rides to reboot scheduling \cite{Panja2021a}; (iii) a worldwide supply chain blockage due to natural disasters \cite{Shughrue2020}, or because of a singular shipping accident [e.g., in the Suez Canal in March 2021 \cite{Lynch2023}]; (iv) a financial crash happening without a compelling fundamental reason and on days without significant news [e.g., the `Black Monday' October 19, 1987 stock market crash \cite{sornette2003}]; or (v) the global financial (and economic) crisis of 2008 that emanated from the US subprime loan market, which represented a small fraction of the US economy, and an even smaller fraction of the global economy \cite{firefighting}. 

\vspace{1mm}
\noindent In view of this, the overarching concept of fragility has been of particular interest for STSs. The concept has emerged in the last 40 years as a generic property of {\it constrained optimization problems\/} or {\it constraint satisfaction problems\/} \cite{Franz2017}, like the well-known Traveling Salesman problem \cite{mezard} or the Max Cut problem \cite{maxcut}, also known as the Spin-Glass problem \cite{mezard} in physics literature. Many problems in STSs are of similar nature and typically exhibit fragility in the sense that they often move towards a true critical point where small perturbations may catastrophically degrade the performance of the system.
\begin{figure*}
\includegraphics[width=\linewidth]{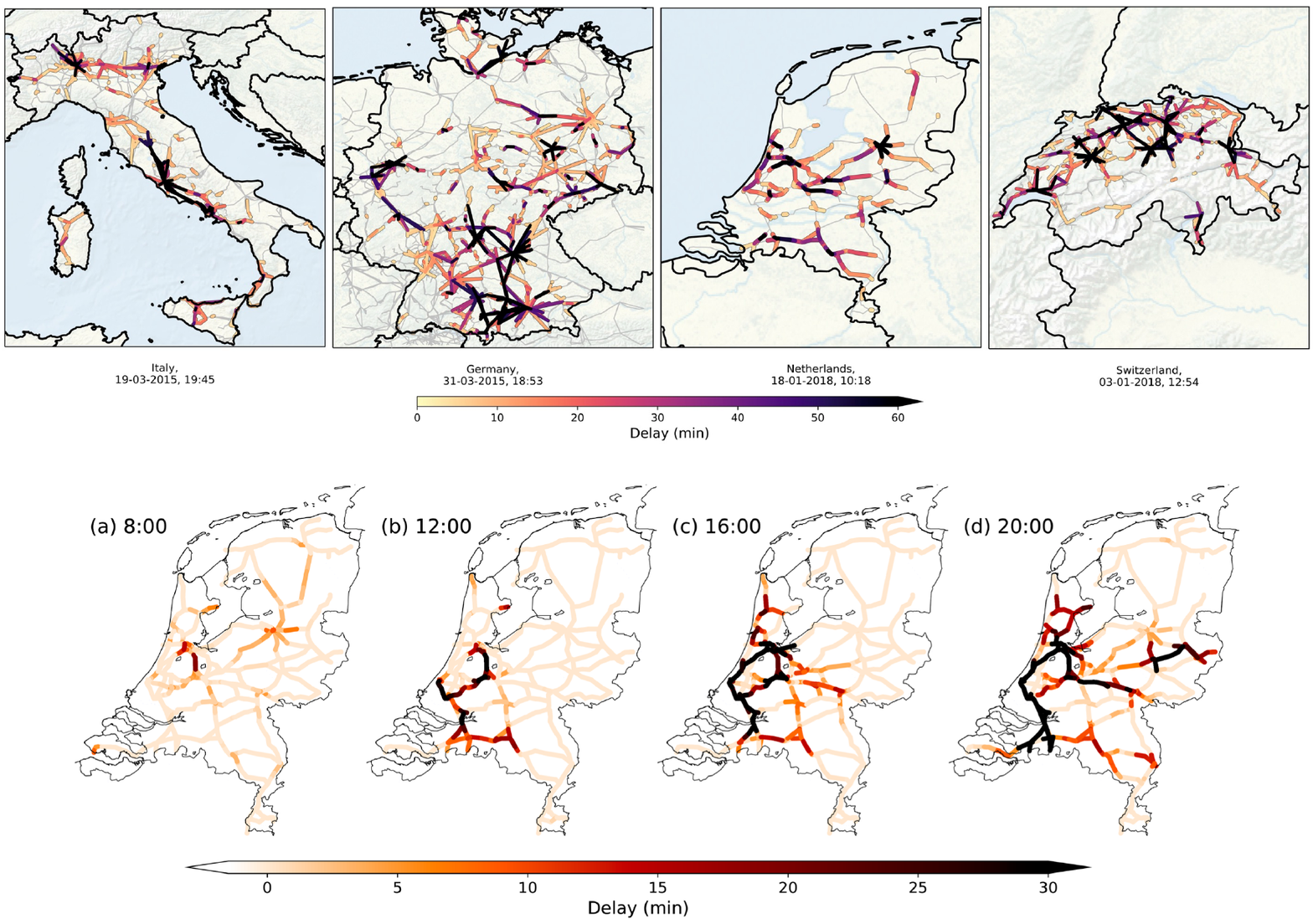}
\caption{Examples of system-wide disruptions in rail transport: Italian, German, Dutch and Swiss Railways (top panel, from left to right). Italy: near-simultaneous occurrence of several problems in the Italian railways in March 2015 --- a major one around Rome, affecting mostly intercity trains, and one between Milan and Venice. Germany: effect of cyclone ‘Niklas’ (31 March 2015) on the German railways. In particular, a specific train near Pegnitz (center-south) was severely damaged by a fallen tree and the rooftop of the Munich station was destroyed, along with multiple smaller incidents across the country. The high risk of more accidents and delays caused the Deutsche Bahn to cancel most of its train activity throughout the day, leaving passengers stranded in major cities like Hannover, Frankfurt, Kassel and Berlin. Netherlands: aftermath of storm ‘Friederike’ in January 2018 in the Netherlands, coinciding with an accident in the north of the country. Fallen trees and damaged overhead lines made the fire department force the Dutch railways to close at multiple stations—resulting in no train activity between the end of the morning and 14:00. Switzerland: in January 2018 (coinciding with storm Burglind/Eleanor in the north-west of Europe) a strong disruption in near Z\"urich (north) rapidly propagated towards the rest of the country. Albeit caused by external triggers such as weather conditions, power outage etc., system-wide disruptions often start localized and small, and systems' internal operations can contribute to the growth of disruptions to system-wide scales, exemplified in the system snapshots of the Dutch railways on February 3, 2012 (bottom panel). On that day winter weather conditions from the night before caused scattered service delays early morning [subpanel (a)]. Subsequently, the delays continued to worsen through the day, ultimately resulting in the loss of overview and an eventual shutdown of a significant part of the system [subpanels (b-d)]. Figures and the caption reproduced from Refs. \cite{Panja2019} and \cite{Panja2021} respectively. \label{disruption_endo}}
\end{figure*}

\vspace{1mm}
\noindent Lately, analyzing STSs, using tools and methods provided by network science, has delivered important insights into such emergent behavior \cite{Rouse2015,Jervis1997}. Examples include the analysis of train operations \cite{Panja2021,Panja2021a} (see Fig.  \ref{disruption_endo}), airline operations \cite{Chung2017}, or the analysis of how firms manage their productions. For the first two cases, development and propagation of delays are archetypal emergent outcomes despite active mitigation measures. For the latter case, the complex interactions among firms trying to manage their productions and inventories, while simultaneously anticipating the actions of their clients and suppliers, can lead to emergent system dynamics \cite{dessertaine2022out}, as demonstrated by the `bullwhip effect', best illustrated through the well-known `beer game'~\cite{Sterman1989}. Since the origin of STSs as a conceptual framework in the 1950s \cite{jaques1951,Emery1965,Baxter2011}, there has been an intense interest in describing and understanding the systems' elements and their interactions, including the transition from one technology or system to another \cite{Nesari2022}. What is largely missing, however, is a focus on {\it time\/} as an operationally critical factor. Only in the last decade has it been realized that disruptions that travel over technological infrastructures or human-technology interaction pathways are aggravated or moderated by their propagation speed, which, given the systems' geographical spread, gives rise to system-specific time scales \cite{Panja2021,Buldyrev2010}. 

\vspace{3mm}
\noindent {\bf Timeliness is a quality standard for socio-technical systems.} A pure focus on system-specific time scales, however, does not underscore the impact of adopting {\it timeliness\/} as a key quality standard for the operations of STSs. Imagine, for example, a local grain shortage even though a sufficient quantity of grain is already purchased and on its way, but simply has not arrived yet at the destination. Indeed, it is often not appreciated that the common element of many socio-technical systems is that their operations are based on a {\it schedule\/}; i.e., on input–output systems or processes that operate {\it on the premise of interdependent events and breaks at specific time intervals\/}. Adherence to the schedule --- system elements being available at the right place at the right time (as scheduled) --- is what we define as timeliness. The schedule provides the reference for timeliness, and delays at specific events are defined with regard to this schedule. In this definition, schedule-based systems do not need to be predetermined, but can evolve dynamically. They also do not need to have a central planner, i.e., can develop decentrally.

\vspace{1mm}
\noindent Smooth functioning of schedule-based systems are synonymous to this definition of timeliness, which has been quietly, ubiquitously and integrally adopted as a key quality standard in most STSs, such as transport systems, supply chain systems, healthcare systems, emergency response systems, computer and web service systems, or agricultural systems. One of the most prolific examples in this context are Toyota's famous `Just-in-Time' and `Kanban' schemes \cite{sugimori1977toyota}, which have been path-breaking innovations for improving efficiency and reducing operational costs. Despite having been widely adopted around the world, these schemes have the potential to lead to major disruptions if an irreplaceable component (e.g., a microchip) lacks a critical quality: being in the factory {\it on time\/}. Buffers of all sorts (for example, with increased inventory, a more diversified supplier base, or alternatives for inputs), which can be abstractified into a {\it temporal buffer\/} incorporated into the schedule, partially mitigate  delays. However, if system operators prioritize cost- and time-efficiencies --- {\it in effect, reducing the temporal buffer\/} --- STSs can become fragile. 

\vspace{1mm}
\noindent Analyzing and understanding this fragility behavior is of paramount importance to manage the trade-off between system efficiency and robustness to perturbation in STSs. The purpose of this perspectives article is to place at the center stage the critical role the concept of timeliness plays in the fragility and operations of STSs, and with it, to emphasize a research agenda that focuses on possible implications, consequences and mitigation measures.

\vspace{3mm}
\noindent {\bf Improving efficiencies of socio-technical systems.} There exists a variety of incentives for STS operators, often reinforced by competitive pressures, to improve cost- and time-efficiencies in order to achieve superior operational results. Here, timeliness as a quality standard plays an important role: {\it improving cost- and time-efficiencies are, in essence, tantamount to reducing temporal buffers while maintaining the scheduled operations\/}. Transport systems provide a direct example of this. Operators in charge of scheduling trains may have the goal to maximize the number of passengers to be transported by the network, i.e., to fit trains into ever tighter schedules. If every train runs as planned, this does indeed result in a larger number of transported passengers. The elements of a train system are however interdependent in their functioning --- a train may not be able to leave a station if another train has not arrived yet, since the crew for the departing train are those from the arriving one --- so tighter schedules can only be implemented by reducing the {\it temporal buffers\/} that are usually kept for allowing the system to absorb possible delays, and/or by having a buffer of replacement crew in place at that station. An operator with an extra eye for cost-efficiency and experiencing well-running train services may, e.g., indeed consider reducing the replacement crew buffer, since the operator must incur costs to maintain this buffer. 
\begin{figure}[!h]
\begin{center}
\includegraphics[width=0.5\linewidth]{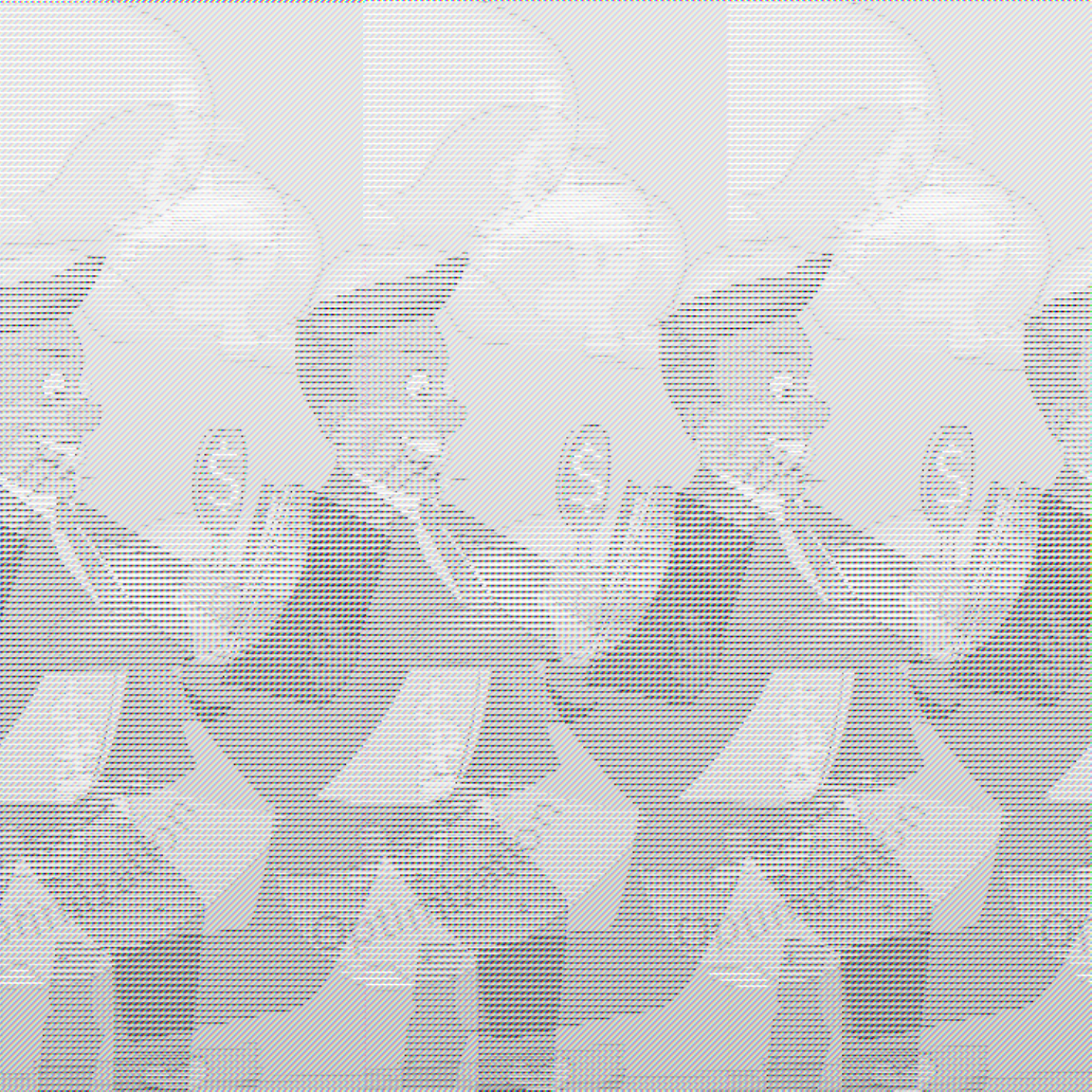}
\caption{A cartoon impression of a STS operator, prompted by incentives (financial or other), attempting to myopically optimizing the system, and in that process, inadvertently pushing the system towards the proverbial cliff edge. \label{cliffedge}}
\end{center}
\end{figure}

\vspace{1mm}
\noindent A similar example exists in just-in-time supply chains. In this case, the temporal buffer analogy is also clear, albeit somewhat more convoluted: to be cost-effective, firms tend to forego inventories of inputs (storage adds to firms' costs) in favor of receiving it at the last minute (Just-in-Time). Hence, such firms may have to stop production if there is any delay at any level of its supply chain. There are, however, also other sources of optimization: consider for example a firm that, in order to obtain better contractual conditions, establishes a contract of exclusivity with one of its suppliers. Such lack of redundancy in the supply chain --- often referred to as substitutability in production networks of firms \cite{carvalho} --- increases the difficulty to find an alternative source for an input that is missing. Firms that keep low inventories or have poorly diversified supply chains lack the appropriate `lead time' --- i.e., {\it temporal buffer\/} --- and are therefore fragile to external shocks~\cite{LafrogneJoussier2022}. The effect of the disruption of Europe's heavily Russia-centered supply of natural gas at the start of the Ukraine war is an example of this at the supranational level \cite{di2022natural}.

\vspace{3mm}
\noindent{\bf Optimizing for efficiencies can make socio-technical systems fragile.} Many optimization problems in STSs belong to the class of constrained optimization: either through the efforts of system operators or by betting on the virtue of an invisible hand, such systems strive to achieve some kind of optimality, with a large amount of constraints. Indeed, optimizing cost and increasing service frequency constrained to reusing the crew from the incoming trains for outgoing trains, as well as input-output production networks --- where each firm attempts to maximize profits by appropriately choosing its suppliers and production technologies to match demand from its customers while minimizing inventories --- are real-world, and complex, examples of constrained optimization problems.

\vspace{1mm}\noindent Such optimization problems typically exhibit fragility in the following sense: the optimal solutions evolve ``chaotically'' with the precise specification of the parameters of the problem. In other words, the optimal solution for one choice of parameters can become sub-optimal, or can even suddenly disappear when these parameters are only slightly changed. Note that this statement can be made precise in the context of spin-glasses, for example, and holds in the limit of a large number of degrees of freedom [for example number of firms, of agents, or of assets, etc.] see, e.g. \cite{mezard}. Such parameter chaoticity is related to the existence of a large number of quasi-optimal solutions, that are nearly equally good in terms of their performance but very different from one another in terms of their realizations (e.g., the actual supplier-client connections in a production network)  \cite{colon}. Moreover, in a large variety of cases, the optimal solution is found to be sitting, proverbially, ``at the edge of a cliff'', in the sense that small perturbations may catastrophically degrade the performance of the system. Such perturbations can either be {\it exogenous\/} \cite{Shughrue2020,Inoue2019}, namely of origin external to the system, or {\it endogenous} \cite{sornette2006endogenous, Wehrli2022}, meaning they originate from within the system. Returning to the train system, a global power failure, or a snowstorm, is clearly an exogenous perturbation, while a mechanical dysfunction of a single train (e.g., due to maintenance issues), or the unavailability of a single crew member, causing a delay, is an endogenous one that may reverberate through the entire system. The boundary between perturbations of exo- vs endogenous origins is, however, not always discernible \cite{Panja2021a}, as seen in Fig. \ref{disruption_endo}.
\begin{figure*}[!h]
\begin{center}
\includegraphics[width=\textwidth]{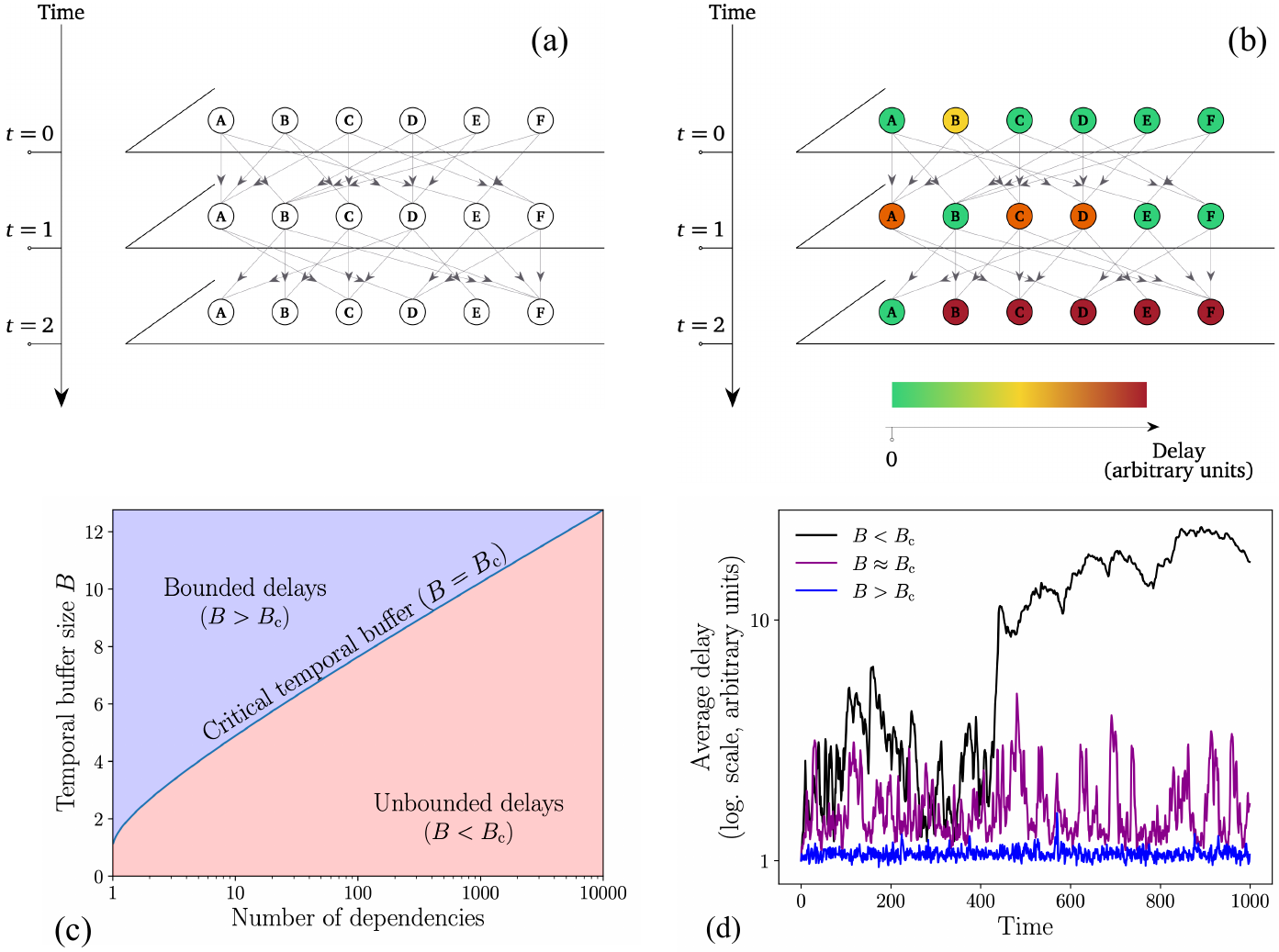}
\caption{A stylized model for the operations of a firm-to-firm production network \cite{Moran2024}. (a) Six firms, A through F, and their operational dependencies across successive temporal layers shown by arrows. (b) An illustration of color-coded production delays propagating downstream in time through this operational dependencies in the network. (c) The criticality curve in the phase diagram of temporal buffer size $B$ and number of dependencies, for the case when each firm has exactly the same total number of operational dependencies. (d) Typical evolution of average delay per firm below, at and above criticality, for the case (c), as a function of time. Note in particular the delay avalanches at $B\approx B_{\text c}$.\label{fig3}}
\end{center}
\end{figure*}

\vspace{1mm}
\noindent In principle, faced with uncertainty about parameter values, system operators should adopt robust strategies. However, as noted above, operators of STSs have a natural tendency to perceive certain safeguards ensuring stability and resilience as unnecessary hindrances to achieving ever higher efficiency. This tendency is particularly strong when the STS has displayed stability for a long time, as overconfidence and/or complacency on the part of system operators make such safety measures or regulations seem unjustified. Driven by the desire to optimize, they come to dismantle these safeguards, inadvertently edging the system critically close to collapse. This is precisely the mechanism proposed by Minsky to explain the recurrence of financial crises \cite{minsky}: regulations limiting the risk of financial institutions are put in place after crashes, thereby lowering the collective perception of risk. Once the memory of the last crisis fades, these safeguards are progressively removed, bringing about another crisis. Such a scenario resembles that of {\it Self-Organized Criticality (SOC)\/}~\cite{bak, bakprl, Baketal93, bouchaud2024self} (for the case of STSs, it is organized by the operators themselves that are part of the systems), illustrated by a sand pile that keeps growing until its slope becomes so steep that an avalanche occurs, (partly) collapsing the sand pile.

\vspace{1mm}
\noindent Moreover, in systems of systems such as a cluster of interdependent production firms, whose individual operations are intricate systems in themselves, it is foreseeable that competitive pressures force individual firms to aggressively optimize for their own cost and time efficiencies. Times during which supply chains run smoothly cause companies to optimize their supply lines by removing redundancies, sourcing raw materials from specific locations and keeping inventories at a minimum. When such myopic optimization is done simultaneously by many firms --- essentially {\it all\/} of them because of competitive pressures --- the entire production system will be pushed to the cliff edge (Fig. \ref{cliffedge}), and will become fragile. Arguably, the more myopic the optimizations, the higher the fragility of the system in its entirety.

\begin{table*}[!h]
\begin{center}
\begin{tabular}{c|c|c|c|c|c|c}
\hline\hline\arrayrulecolor{white} series&cadence&epoch&$n$&growth rate&rMSR&Kurtosis\\&&&&[yr$^{-1}$]&[yr$^{-1}$]&[dimensionless]\\\arrayrulecolor{black}\hline GDP & annual & 1950-2022 & 72 & 0.0681 & 0.0402&0.3293\\ GDP & quarterly & 1995-2022 & 112 & 0.0343 & 0.0153&2.332\\ Imports & annual & 1950-2022 & 72 & 0.0754 & 0.0744 & 0.7229\\Imports & monthly & 2015-2022 & 96 & 0.0554 & 0.0463 & 0.1990\\\hline
\end{tabular}
\end{center}
\caption{Statistical measures of period-on-period volatility and their root mean square residues in the National Accounts data for the Netherlands. Noteworthy is that the rMSR values are of the same order of magnitude as the growth rate, while the Kurtosis values indicate that the volatility distribution is likely fat-tailed. This is remarkable, because the measures are already highly aggregated from many underlying time series and would be expected to have very close to normally distributed properties. Data source: Statistics Netherlands (opendata.cbs.nl).\label{tab1}}
\end{table*}

\vspace{3mm}
\noindent {\bf The limiting operational performance of timeliness as critical fragility.} Given our argument earlier --- {\it viz.\/}, improving cost- and time-efficiencies are, in essence, tantamount to reducing temporal buffers while maintaining scheduled operations --- when we bring {\it timeliness\/}, the universally-adopted quality standard for STSs, into the picture, we expect the proximity to criticality of constrained optimized solutions to manifest itself in terms of the system's propensity to accumulate and propagate of {\it operational delays\/} that may run out-of-control. Indeed, this is an overarching aspect of STSs ranging from financial markets (e.g., liquidity not available when needed), to transport (e.g., physical trains/planes or crew members not present at the start of a scheduled service), to production (e.g., raw materials not available at firms when goods need to be produced), to food systems (e.g., grain has not arrived in time at the port for shipping). Specifically, from statistical physics, we expect the systems to be extremely sensitive to delays close to criticality --- the closer the system is to the cliff edge, the higher the magnitudes of the delays, and the longer their persistence. We foresee such behavior to be characterized as {\it critical behavior in the temporal dimension\/}.

\vspace{1mm}
\noindent These ideas have been given a precise form by means of a stylized model, analysed in detail in Ref. \cite{Moran2024}, and here we only summarize the main results.  The model describes the propagation of delays across a (spatio-)temporal network of tasks, mimicking the geographical spread and temporal ordering of planned operational interactions among the components of STSs (Fig. \ref{fig3}, for a firm-to-firm production network). The delay at a certain node of the temporal network is given by a random variable (modeling unexpected events specific to that node) plus the {\it maximum\/} delay of its incoming nodes, from which is subtracted a temporal buffer value $B$. When the second term is negative, the temporal buffer suffices to absorb all accumulated delays at that temporal node, and this contribution is set to zero. This simple model displays precisely the type of behavior described above: as soon as nodes become interdependent, there exists a critical temporal buffer value $B_{\text c}$ below which delays keep accumulating, leading to a system wide crisis. Just above $B_{\text c}$, {\it delay avalanches\/} of all sizes appear spontaneously, with no relation to the amplitude of the initial perturbations [see Extended Data Fig. 1 and the associated Methods text from Ref. \cite{Moran2024}]. Since maintaining buffers entails costs for the operators of STSs, one expects the systems to operate close to $B_{\text c}$, and therefore to be generically fragile, and any system that operates on the basis of a schedule will show a phase transition similar to Fig. \ref{fig3}(c-d). This is indeed a crucial aspect of input-output networks, which, in order to avoid uncertainties related to production delays, must rewire when some firms go bankrupt or cease to function, like what happened in the Fukushima region after the 2011 tsunami \cite{fukushima}. 

\vspace{3mm}
\noindent{\bf Volatility as a possible consequence of critical fragility.} Both financial markets and large economies exhibit significant volatility (see Table \ref{tab1} for the Dutch economy), on a much greater scale than what traditional equilibrium models would predict. This phenomenon, often called the ``excess volatility puzzle'', or the ``small shocks, large business cycle puzzle'' \cite{Bernanke1996,firefighting}, shows a gap in these models: they typically assume perfect agent coordination and neglect the effects of supply-demand imbalances. The role of the production network in amplifying fluctuations is indeed studied, but only through the properties of its stationary equilibria --- see Ref.~\cite{carvalho} for a recent review. Other approaches, such as the study of the optimal scheduling work done by~\cite{Golub2009}, integrate dynamics and study the problem of an individual firm faced with delays in its operations. Such approaches are, however, still done by assuming stationarity and are seldom considered at a macroeconomic scale.

\vspace{1mm}
\noindent Other scenarios have been studied through agent-based modeling to account for the dynamic nature of economic systems. In these models, inventories serve as buffers that absorb fluctuations~\cite{Hallegatte2013,Inoue2019, Colon2020, dessertaine2022out}. However, maintaining large inventories incurs costs, and thus firms have a natural incentive to minimize them, potentially exacerbating instability.  This points to a link with self-organized criticality for economic systems, as first proposed by late Per Bak and his coauthors in 1993 \cite{bak}, and recently revisited in Refs. \cite{moran, nirei2024repricing,bouchaud2024self}.

\vspace{1mm}
\noindent More work, both theoretical (on realistic models of critical fragility in supply chains as limiting operational performance of timeliness) and empirical (looking for signatures of critical fragility in production data), is obviously needed to affirm or contradict our posit. We believe that quantifying the role of timeliness is a very important research question, both in the economic context and more generally, in the overarching context of STSs. 

\vspace{3mm}
\noindent{\bf Concluding remarks.} Until recently the well-functioning of supply chains may have caused managers to overestimate the precision of their estimates of disruptions and ignore tail risks. Covid-19 and the Ukraine war caused firms to adopt measures (near-shoring, diversification of suppliers) that may mitigate some threats to timeliness. Adding such resilience will increase costs and degrade strict economic performance [as is also the case in a theoretical equilibrium model \cite{kopytov2024}], but may keep the solution at a safe distance away from the cliff edge. However, there is no guarantee that once we have another long period of calm, firms would not, once more, adopt behavior that would bring the system to critical states. Elucidating the very mechanisms leading to instabilities, failures and system-wide crises in socio-technical systems is, therefore, crucial to finding remedies, mitigation measures and proposing adequate regulations. In fact, there is still a lot of research needed to find ``anti-fragile'' solutions \cite{taleb, Hynes2022} for systems to spontaneously improve when buffeted by large shocks.

\vspace{3mm}
\noindent {\bf Author contributions.} J-PB, JM and DP analyzed critical fragility. All authors contributed to the manuscript.

\vspace{3mm}
\noindent{\bf Competing interests.} The authors have no competing interests to declare.

\vspace{3mm}
\noindent \textbf{Code availability.} Data for Fig. \ref{disruption_endo} have been published previously (by one of the authors) and are already publicly available for free. Code to perform the simulations in Fig. \ref{fig3} can be found at \url{https://github.com/jose-moran/timeliness_criticality}.

\section*{References}



\end{document}